\documentclass[a4paper,fleqn,useAMS,usenatbib]{mnras}

\pdfoutput=1

\usepackage{mathptmx}

\usepackage[T1]{fontenc}
\usepackage{ae,aecompl}

\usepackage[dvips]{graphicx}
\usepackage{amsmath}
\usepackage{amsfonts}
\usepackage{amssymb}
\usepackage{comment}
\usepackage[dvipsnames]{xcolor}
\usepackage[%
        ]{hyperref}
        
\hypersetup{
	colorlinks=true,	
	urlcolor=MidnightBlue,		
	pdfpagelabels=true,
	hypertexnames=true,
	plainpages=false,
	naturalnames=true,
	pdftitle={The giant nature of WD\,1856\,b},    
	pdfauthor={Kipping},     
	linkcolor=WildStrawberry,          
	citecolor=ForestGreen,        
}

\newcommand{\multi}{{\sc MultiNest}}
\newcommand{\cofiam}{{\sc CoFiAM}}

\newcommand{\kepler}{{\it Kepler}}
\newcommand{\tess}{{\it TESS}}
\newcommand{\gaia}{{\it Gaia}}

\newcommand{\pdf}{ \mathrm{Pr} }

\newcommand{\forecaster}{{\tt forecaster}}
\newcommand{\wwwcoolworlds}{\href{https://github.com/davidkipping/WD1856/}{this URL}}

\title[The giant nature of WD\,1856\,b]{The giant nature of WD\,1856\,b implies that transiting rocky planets are rare around white dwarfs}
\author[Kipping]{David Kipping$^{1}$\thanks{E-mail:
\href{mailto:dkipping@astro.columbia.edu}{dkipping@astro.columbia.edu}}\\
$^{1}$Dept. of Astronomy, Columbia University, 550 W 120th Street, New York NY 10027}

\date{Accepted 2023 October 20. Received 2023 September 27; in original form 2023 August 31}

\pubyear{2023}

\begin{document}
\label{firstpage}
\pagerange{\pageref{firstpage}--\pageref{lastpage}}
\maketitle

\begin{abstract}
White dwarfs (WDs) have roughly Earth-sized radii - a fact long recognized to facilitate the potential discovery of sub-Earth sized planets via transits, as well atmospheric characterization including biosignatures. Despite this, the first (and still only) transiting planet discovered in 2020 was a roughly Jupiter-sized world, found using \tess\ photometry. Given the relative paucity of giant planets compared to terrestrials indicated by both exoplanet demographics and theoretical simulations (a ``bottom-heavy'' radius distribution), this is perhaps somewhat surprising. Here, we quantify the surprisingness of this fact accounting for geometric bias and detection bias assuming 1) a bottom-heavy \kepler\ derived radius distribution, and 2) a top-heavy radial velocity inspired radius distribution. Both are concerning, with the latter implying rocky planets are highly unusual and the former implying WD\,1856\,b would have to be highly surprising event at the $<0.5$\% level. Using an HBM, we infer the implied power-law radius distribution conditioned upon WD\,1856\,b and arrive at a top-heavy distribution, such that $0.1$-$2$\,$R_{\oplus}$ planets are an order-of-magnitude less common than $2$-$20$\,$R_{\oplus}$ planets in the period range of 0.1-10\,days. The implied hypothesis is that transiting WD rocky planets are rare. We discuss ways to reconcile this with other evidence for minor bodies around WDs, and ultimately argue that it should be easily testable.
\end{abstract}


\section{Introduction}
\label{sec:intro}

WD\,1856\,b (more formally WD\,1856+534\,b) is the first transiting exoplanet
discovered around a white dwarf (WD) star; indeed it remains the only such
world even three years after its announcement \citep{vanderburg:2020}. In many
ways, the detection of such a planet was highly anticipated, with indirect
evidence for their existence having been amassing for years prior to
WD\,1856\,b. For example, 25-50\% of WDs exhibit the spectral fingerprints of
metal contamination \citep{koester:2014}, indicative of in-falling planetary
debris \citep{jura:2014,bonsor:2020}. Such debris was even observed in transit
for WD\,1145+017 \citep{vanderburg:2015}.

Encouraged by these early signs, searches for WD exoplanets had been attempted
in previous years \citep{fulton:2014,vansluijs:2018}. Fueling these searches was
the potential opportunity to spectroscopically characterize the atmosphere of
temperate, rocky planets around such stars using existing facilities to
exquisite precision \citep{limbach:2022}. Indeed, Earth-like planets around
WDs represent the only class of planet for which biosignatures could be
detected with existing facilities, namely JWST \citep{kaltenegger:2020}.
Although planets of any size around WDs provide important clues to post
main sequence evolution \citep{veras:2021}, the potential for observational
evidence of life - arguably the most impactful possible scientific discovery
- with existing facilities is completely unique and thus has an outsized
influence in these discussions.

Although WDs might seem a strange place to expect life, the potential
for a long-lived habitable zone has been long recognized \citep{agol:2011},
at approximately 1/100th of the Earth's semi-major axis (periods ${\sim}$day).
The recognition of their unique observational advantages has fuelled enormous
interest in the possibility of Earth-like WD exoplanets. For example, there is
now interest to searching for such planets using LISA \citep{tamanini:2019,
danielski:2019,kang:2021}, \gaia\ \citep{sanderson:2022}, CHEOPS
\citep{morris:2021}, LSST \citep{cortes:2019}, DECam \citep{dame:2019},
proposed UV missions \citep{fleming:2020,shvartzvald:2023} and JWST Cycle 2
observations (IDs 3621, 3652, 4403).

Despite this excitement, there are perhaps some reasons for concern hinted at
in the very discovery of WD\,1856\,b. Specifically, it is a \textit{giant}
exoplanet with a radius of $(10.4\pm1.0)$\,$R_{\oplus}$
\citep{vanderburg:2020}\footnote{Assuming what is very likely a circular orbit
given the compact orbital period of $1.4$\,days.}, which is argued in what
follows to be intuitionally surprising.

First, it appears unlikely that WD\,1856\,b is a so-called ``second-generation''
planet forming from a WD disk post-MS (main sequence). \citet{lieshout:2018}
consider this formation channel and find that minor planets, up to potentially
Earth-mass, are plausible - but a $\gtrsim 100$\,$M_{\oplus}$ is certainly not
anticipated. Accordingly, WD\,1856\,b likely formed along soon after the star
itself formed, and subsequently survived the RGB and AGB phases
\citep{vanderburg:2020}. In this way, the planet's current orbital period of
1.4\,days is surely not the MS orbit, since this would be deep inside the
star. Instead, the planet likely lay beyond $\gtrsim$\,AU and migrated inwards
via planet-planet scattering \citep{nordhaus:2013,mustill:2018}, Kozai-Lidov
oscillations \citep{stephan:2017} or even common-envelope evolution
\citep{lagos:2021}.

With this point established, it's not constructive to compare the radius-period
\textit{tuple} of WD\,1856\,b with demographics of MS stars. However, assuming
that the planet's current radius is approximately representative of its MS
radius, then it is useful to compare its radius (in isolation) to the exoplanet
demographics around FGK stars. A common result from exoplanet demographics
studies is that Jupiter-sized/mass planets are less common than their
sub-Jovian counterparts (e.g. \citealt{howard:2012}, \citealt{wittenmyer:2016},
\citealt{he:2019}), a result also supported by theoretical simulations of
planet formation \citep{mordasini:2012}. Despite this, one might argue
that WD\,1856\,b's giant radius is not surprising though because giant
planets typically have much deeper transits, as well as longer durations
and enhanced transit probabilities \citep{beatty:2008}. However, the roughly
Earth-like radius of the host star here certainly erodes away the first
(and most major) of those advantages, and thus at least intuitionally
\textit{there is} something surprising in that the first WD transiting exoplanet
is a giant planet.

This argument is qualitative and hand-wavy as presented thus far, but ultimately
motivated us to investigate this question more rigorously. In this work, such a
calculation is performed where the surprisingness of WD\,1856\,b's giant radius
is quantified in Section~\ref{sec:pvalue}. Following this, we ask what kind of
radius distribution is required to dissolve this tension using a Hierarchical
Bayesian Model (HBM) in Section~\ref{sec:hbm}. However, both of these sections
requires a completeness model for \tess, which is introduce first in 
Section~\ref{sec:completeness}. Although a single data point, WD\,1856\,b in
this case, has limited power to make claims about an entire population, we
discuss the provocative implications of our work in
Section~\ref{sec:discussion}, which fortunately will be testable in coming
years.

\newpage
\section{Completeness Model for WD\,1856}
\label{sec:completeness}

\subsection{Overview}
\label{sub:completenessoverview}

A basic component in considering the likelihood of detecting a planet such as
WD\,1856\,b is the detection probability, also known as the sensitivity,
true positive rate or completeness. \citet{vanderburg:2020} report that
WD\,1856\,b was initially detected in just a single sector of \tess\ data
(Sector 14) and thus in this work we strictly consider what other exoplanet
radii and orbits would be equally detectable in that sector. However, this
requires us to determine some function which evaluates completeness given an
input set of planet parameters.

The basic starting point in this discussion is signal-to-noise ratio (SNR),
assumed in what follows to be the cumulative SNR of multiple transits within
a given time frame. We expect the true positive rate to directly map onto SNR,
but before we discuss this mapping procedure, we first need to define SNR.
Given the grazing nature of WD\,1856\,b, a box-like transit assumption is a
poor approximation and thus so too are traditional transit SNR formulae of the
form $\delta/\sqrt{W}$, where $\delta$ is the transit depth and $W$ is the full
width half maximum duration. This issue is explored in detail in
\citet{kipping:2023}, where it is shown that if limb darkening is
non-negligible (as is true in \tess\ bandpass), the numerical integration
is the only path to accurately computing the SNR.

Accordingly, our first task is to generate an efficient algorithm into which
we can take in a set of planet input parameters and quickly return
the corresponding SNR - assuming the same photometric precision and data
volume as measured in \tess\ Sector 14 for WD\,1856.

\subsection{Computing SNRs}
\label{sub:computingSNRs}

To compute the transit SNR of a given simulated transit we follow a series
of steps. First, given a set of input parameters, a \citet{mandel:2002} light
curve is generated for a single transit from $-0.025$\,days to $+0.025$\,days
(which if sufficient to span even our longest period planets on equatorial
transits) using a cadence of one hundredth the integration time, which itself
is set equal to that of \tess's Sector 14 data (2\,minutes). Integration time
effects are handled using numerical resampling following \citet{binning:2010}
with $N_{\mathrm{resamp.}}=30$. We use a quadratic limb darkening law,
interpolating the \citet{claret:2020} WD DA $Z=0$ table of coefficients
(Table 100C) for $\log g = 7.915$ and $T_{\mathrm{eff}} = 4710$ (best reported
values from \citealt{vanderburg:2020}) to give $u_1 = 0.0101$ and
$u_2 = 0.4091$.

Next, the simulated light curve is adjusted for blended light, following
the prescription of \citet{nightside:2010}. However, the challenge here is that
the blend factor was never reported in \citet{vanderburg:2020}, as their
final set of parameters instead using photometry from GTC and \textit{Spitzer}.
These were obtained with much tighter angular apertures for which contamination
was simply not an issue. If we similarly assume the GTC photometry is
unblended\footnote{The \textit{Spitzer} data is of poorer quality and features
challenging red noise components and thus was not used here.}, we
can jointly fit the \tess\ photometry\footnote{Including all 13 sectors for
which WD\,1856 had been observed at the time of writing.} with the GTC
photometry to infer the most probable blend factor of each sector, assuming the
radius of the planet is approximately achromatic.

We thus obtained the corrected GTC photometry (A. Vanderburg, priv. comm.),
detrended the \tess\ photometry ourselves using \cofiam\ \citep{kipping:2012},
and then jointly fit both data sets using a \citet{mandel:2002} light curve
model and \multi\ \citep{feroz:2009}. The maximum \textit{a-posteriori} blend
factor for Sector 14 was determined to be $(F_{\mathrm{WD\,1856}} +
\sum F_{\mathrm{blend},i})/F_{\mathrm{WD\,1856}} = 9.580$. With this blend
factor, we can now correct the simulated light curves to the correct depths
that \tess\ would observe.

Next, the light curve is spline interpolated and the interpolation function is
numerically integrated to compute the SNR of a single transit following
Equation~(15) of \citet{kipping:2023}. This formula requires inputting a
noise value as computed over the same time unit basis as the photometry (in
our case days), $\sigma$. Hence, $\sigma$ will equal the typical measurement
uncertainty of the \tess\ Sector 14 WD\,1856 photometry divided by the
square root of the number of points per day, or simply $\sigma =
\sigma_1\sqrt{\mathrm{cadence}}$. In our case, we set $\sigma_1$ to be
the median uncertainty of the whole sector's set of reported 
uncertainties.

To account for multiple transits, we multiply the single-transit SNR by
$\sqrt{N}$, where $N$ is the number of epochs occurring within Sector 14,
calculated using the simulated ephemeris parameters. To account for
data gaps, we only count epochs for which the expected time of transit
minimum has a nearest Sector 14 data point within one cadence. The final
SNR is then stored to file.

In practice, the above was too computationally expensive to practically
integrate into a Bayesian inference model with large numbers of calls, as used
later in Section~\ref{sec:hbm}. We thus decided the best approach was to
calculate a grid of SNRs based on a broad range of input parameters, and then
use this grid to train an interpolation model that would be far faster to
evaluate than the SNR numerical integration each time.

Accordingly, the SNR calculation was repeated across a grid of input
ratio-of-radii, $p$, impact parameters, $b$, and orbital periods, $P$.
We define a log-uniform grid in $p$ such that the minimum value
corresponds to $0.1$\,$R_{\oplus}$ and the maximum is $20$\,$R_{\oplus}$
(assuming $R_{\star}=0.0131$\,$R_{\odot}$; \citealt{vanderburg:2020}).
The impact parameter grid is also log-uniform from $0.01$ to
$(1+p_{\mathrm{max}})$. Finally, the orbital period grid is log-uniform
from $0.1$ to $10$\,days. We carefully selected the grid to be
approximately cubical but reach ${\sim}100,000$ grid points of
non-zero SNR (which was sufficiently dense but not computationally
prohibitive). We elected to use $54$ grid points for $p$ and $b$, and
$47$ for $P$, giving 134,514 grid positions but only 103,447 led to
non-zero SNR (corresponding to non-transiting geometries).

\subsection{SNR of WD\,1856\,b}
\label{sub:1856SNR}

At this point, it's worth discussing the SNR of WD\,1856\,b specifically.
From \citet{vanderburg:2020}, we know that the \tess\ SPOC pipeline
identified WD\,1856\,b from Sector 14. \citet{vanderburg:2020} do not
report the Sector 14 \tess\ SNR, but the ExoFOP DV report
states a multi-event statistic (a proxy for SNR) of MES$=12.1$.

Let us now compare this to value calculated using the same sector but with
the \citet{kipping:2023} SNR definition. That work presents both a discrete
and continuous version of the SNR expression; the former is intended for real
data and the latter for predicting the SNR of model functions. Starting with
the discrete version, we can calculate the $\Delta\chi^2$ between the
best-fitting transit model (taken from our earlier light curve fits in
Section~\ref{sub:computingSNRs}) and a null (flat-line) model on a
point-by-point basis, which gives SNR$=18.80$ for Sector 14.

Alternatively, we can interpolate the best-fitting model and then numerically
integrate the function setting the noise to be equal to the sector-median
\tess\ reported uncertainties, which gives SNR=$19.51$ for Sector 14. The
value from the continuous method is slightly different to the discrete
version, but still within 10\%. This happens because the latter tracks the
actual observed noise behaviour whereas the former assumes expectation value
Gaussian statistics (i.e. averaged over many experiments versus a single
stochastic realization).

In any case, both theoretical SNRs are ${\sim}$60\% higher than that
reported in the \tess\ Sector 14 DV report. This may be due to a possible
difference in how SNR is defined. In particular, the \citet{kipping:2023}
is in principle the maximum possible value under all definitions and
obviously post-dates the \tess\ pipeline development. Further, our SNR values
benefit from a refined best-fit model thanks to the GTC data and 12 other
\tess\ sectors, will inevitably leads to a differences between the
best-fitting model we use to calculate SNR and that used by the \tess\
pipeline using Sector 14 data alone. Recall - SNR is an act of model
comparison and thus always requires a formal model.

Perhaps the most likely explanation is simply that the SPOC pipeline is
optimized for much longer transits than that of WD\,1856\,b and thus
plausibly underestimates the significance of transits like this. In
any case, the SNR is certainly sufficiently high that we should expect
that the completeness probability was very high, as discussed further
in the next subsection.

\subsection{Completeness Curve}
\label{sub:completeness}

A basic question is - how should we map the true-positive probability (TPP),
often called the completeness, to SNR? We should expect TPP to be a
monotonically increased function with respect to SNR, with boundary conditions
of TPP$=0$ at SNR$=0$ and TPP$=1$ at SNR$=\infty$. This actual function that
represents this curve really comes down to the detailed mechanics of how the
\tess\ pipeline identifies, vets and flags \tess\ Objects of Interest (TOIs).

The precise shape of this function was a subject of intense discussion and
iteration in the previous \textit{Kepler Mission}, since it strongly affects
occurrence rate calculations. For example, \citet{howard:2012} attempted the
first such occurrence rate calculation prior to any effort to measure this
completeness curve and thus made the simplifying assumption of a Heaviside
Theta function with a trigger value at SNR$=10$. This was later replaced
with the linear ramp of \citet{fressin:2013}. Later, \citet{christiansen:2016},
presented a calibrated curve that resembled a logistic function and equally the
cumulative density function of a gamma distribution. However, it should be
noted that it was never discovered why these functions truly worked from a
first principles perspective (J. Christiansen, priv. comm.), nor whether the
trained input parameters would be in anyway generalizeable to future missions
like \tess.

Unfortunately, to our knowledge, there is no published \tess\ general-purpose
completeness curve. Attempting to write an \tess\ pipleine emulator and
calibrate its completeness function is a formidable task far beyond the scope
of this work. Instead, for the sake of simplicity, we will assume that the
completeness function is a Heaviside Theta function with a trigger value at
SNR$=10$. This is the same assumption made by \citet{howard:2012} working with
\kepler\ data prior to the any \kepler\ completeness maps being published. We
justify this choice that the \tess\ pipeline uses the same MES (multiple event
statistic - a proxy to SNR) threshold as \kepler\ of 7.1 to flag candidates
\citep{guerrero:2021}, and for \kepler\ at least the completeness function
reaches 90\% beyond 10 \citep{christiansen:2016}. Whilst far from perfect, we
consider this the most pragmatic approach in the current situation.

\subsection{Interpolating Detection Probabilities}
\label{sub:contours}

Equipped with our grid of 103,447 training SNRs, we are ready to define an
interpolation function. We initially tried out-of-the-box interpolators and
machine learning packages but found that they did not reliably reproduce
the expected behaviour smoothly and so took a slightly different approach.

We begin by extracting a single slice of constant orbital period from our
training data and filling out any missing grid positions to create a
regular $\{p,b\} \to \mathrm{SNR}$ grid mapping. We train use a Hermite
cubic interpolator on this grid. Using this function, we input $b$ as its
first unique grid value and then freely vary $p$ until we obtain
$\mathrm{SNR}=10$ (our critical threshold for a detection). This $p$ value
is saved as the corresponding critical $p$, $p_{\mathrm{crit}}$, and then
we repeat for all $b$ grid positions. Near the edges of the grid, we switch to
2nd and 1st order Hermite interpolation as appropriate. The resulting list of
$\{p_{\mathrm{crit}},b\}$ positions define the iso-SNR10 contour for this
particular choice of $P$. We thus repeat for all $P$ grid positions to
create 47 such contours.

For a fixed choice of $b$, we expect $p_{\mathrm{crit}}$ to monotonically
increase with $P$. We largely see this behaviour but the contours get very
close together so to ensure this behaviour against numerical error we
regressed quadratic functions in each $b$-slice and use these predictions
to slightly refine the contours. The final contours are then saved to file
and used in what follows.

\newpage
\section{The Surprisingness of WD\,1856\,b}
\label{sec:pvalue}

\subsection{Simulating a Catalog of Detected WD Transiting Exoplanets}

Armed with the completeness model from Section~\ref{sec:completeness}, we are
now well-positioned to produce a simulated catalog of detected WD transiting
exoplanets. This is accomplished with three ingredients. First, we need to
simulate a population of WD exoplanets with random but representative
physical/orbital parameters. Second, we need down-sample to only
those which would transit WD\,1856. Third, we need to down-sample further
to only those which would be detectable using one sector of \tess\ data,
as occurred for WD\,1856\,b \citep{vanderburg:2020}.

Step three has already been solved for in Section~\ref{sec:completeness}. Step
two is trivially through simple geometry. In our implementation, a simulated
planet has assigned orbital period, which is converted into $a/R_{\star}$
via Kepler's Third Law and assuming $M_P \ll M_{\star}$. The planet is then
assigned a uniform random impact parameter between 0 and $a/R_{\star}$,
corresponding to a uniform distribution in the cosine of inclination angle
(i.e. a isotropic distribution). Only planets for which $b<(1+R_P/R_{\star})$
are flagged with a successful transit flag. We prefer to use this brute force
approach rather than generating a random Bernoulli variable with the transit
probability because it allows us to assign and track and actual $b$ value to
each planet which is then consistently used in the step three completeness
calculation.

This just leaves us with step one, generating the actual tuples of
period-radius. In what follows, we only consider periods between 0.1 and
10\,days, since interior to this tidal destruction is almost guarenteed
and exterior the transit probability drops off precipitously. The period
distribution of WD exoplanets within this range is wholly unknown at this point.
However, \kepler\ statistics broadly resemble a log-uniform distribution
\citep{dfm:2014}, especially in fairly localized windows of period space like
this. Further, the period distribution is really a nuisance parameter in this
work and our central interest is the marginalized result for the radius in any
case.

Proceeding with this period distribution, we now just need a radius
distribution. Unfortunately, there is no observed inference of the radius
distribution of $1-10$\,AU exoplanets. We basically have two possible options.
First, we can take the well-measured radius distribution of $<$\,AU exoplanets
(specifically using \kepler\ data) and extrapolate it out to to longer periods.
Or, alternatively, we could take the mass distribution of 1-10\,AU giant
planets found from radial velocity surveys and extrapolate it down to smaller
masses (and also convert masses to radii). In truth, neither is likely to
resemble reality but they do produce radically different radii distributions
which at least serve as mirror hypotheses to compare to. A third, and arguably
least worst option, is to use neither forward model and instead just simply
attempt to infer the radius distribution from the single datum of WD\,1856\,b,
which is precisely what we will do in Section~\ref{sec:hbm}. But we proceed
with these two forward models in what follows to at least highlight the
surprisingness of WD\,1856\,b conditioned upon them, however questionable
their extrapolated models may be.

\subsection{A Bottom-Heavy Radius Distribution: Extrapolating \kepler\ Statistics to $>$\,AU}

Power-law radius distributions have been a common approach for modelling the
\kepler\ exoplanet radius distribution since the first demographics papers
(e.g. \citealt{howard:2012}). In recent years, these models have typically been
modified to a broken power-law structure, with an inflection point, $R_C$,
occuring around the mini-Neptune regime (e.g. \citealt{burke:2015}), which
better matches the data. In this work, we elected to use the broken power-law
of \citet{he:2019}, which has an inflection point at $R_C = 3$\,$R_{\oplus}$. In
that work, several models are presented and we use the ``non-clustered''
version - given that exoplanet multiplicity is yet to be demonstrated for WDs.
This model is calibrated against the \kepler\ catalog using orbital periods
of 3 to 300\,days. The 300\,day cap exemplifies why this \kepler-based model
formally says nothing about $>$\,AU exoplanets, and thus we are forced to
assume it holds in this longer period regime as an extrapolation. Formally,
our radius distribution is

\begin{equation}
\pdf(R_P) =
\begin{cases}
k \frac{R_C^{-\alpha_2}}{R_C^{-\alpha_1}} R_P^{-\alpha_1} & \text{if } R_{\mathrm{min}} < R_P \leq R_C ,\\
k R_P^{-\alpha_2} & \text{if } R_C < R_P \leq R_{\mathrm{max}} ,\\
0 & \text{otherwise},
\end{cases}
\end{equation}

where $k$ is a normalization constant set such that the sum of
probabilities equals unity. For $\alpha_1$ and $\alpha_2$, each simulation
draws a random realization for them based on the reported values by
\citet{he:2019}. Specifically, we draw from asymmetric normals
using\footnote{Note the signs are reversed to that of \citet{he:2019} due to
the minus sign in our power-law definition.} $\alpha_1 = 1.08_{-0.20}^{+0.31}$
and $\alpha_2 = -5.30_{-0.40}^{+0.53}$. In this model (and all other models),
the minimum and maximum allowed radius is set to $0.1$\,$R_{\oplus}$ and
$20$\,$R_{\oplus}$.

For this model, and indeed the others that follows, we apply a final step
of planet elimination due to tidal destruction. We remove all simulated
planets that orbit within $R_{\mathrm{Roche}} = 2.44
(\rho_{\star}/\rho_P)^{1/3}$, where for $\rho_{\star}$ we adopt the value
of \citet{vanderburg:2020}. Since we work in period space, not semi-major
space, we convert between them using Kepler's Third Law. Planetary
densities are estimated using \forecaster\ \citep{chen:2020} set in
deterministic mode. Above a Saturn-radius, the mass-radius relation is
nearly flat (and thus degenerate) and so we simply assume all planets
in this regime are approximately Saturn-mass.

\subsection{A Top-Heavy Radius Distribution: Extrapolating \textit{CLS} Statistics to Sub-Neptunes}

\kepler\ inferences benefit from the enormous statistical power of the
\textit{Kepler Mission} - a mission designed specifically for this purpose
\citep{borucki:2003}. However, like all transit surveys it suffers from a
strong bias towards inner orbits and thus has poor coverage beyond an AU. Since
WD\,1856\,b likely originated from beyond this distance, it's unclear that
extrapolating the $<1$\,AU \kepler\ radius distribution is representative of
where this planet spent most of its life during the MS phase. Accordingly, we
need a radius distribution for planets at wider orbits. Since only transits
provide radius (semi-)directly, other methods sensitive to $>$\,AU will
unfortunately only provide planetary mass. Nevertheless, we can forecast the
corresponding radii using a mass-radius relation \citep{chen:2017}.

For our mass distribution, we take the results from the California Legacy
Survey (CLS; \citealt{howard:2010}) presented in \citet{fulton:2021}. We
start by digitizing their Figure~6, which shows the binned occurrence rate
versus exoplanet mass in the range 1-5\,AU; a highly plausible initial
pre-WD position for WD\,1856\,b. Unfortunately, the distribution only extends
down to 30\,$M_{\oplus}$, since radial velocities are presently unable
to detect $>$AU terrestrial mass planets. As a result, we will be forced to
extrapolate this distribution in mass space to address the central question
of this work.

To make progress, we proposed an asymmetric normal distribution as the
underlying unbinned occurrence rate distribution (in log-mass space),
characterized by a mean, $\mu$, a negative standard deviation,
$\sigma_{-}$, a positive standard deviation, $\sigma_{+}$.

\begin{figure}
\begin{center}
\includegraphics[width=8.4cm,angle=0,clip=true]{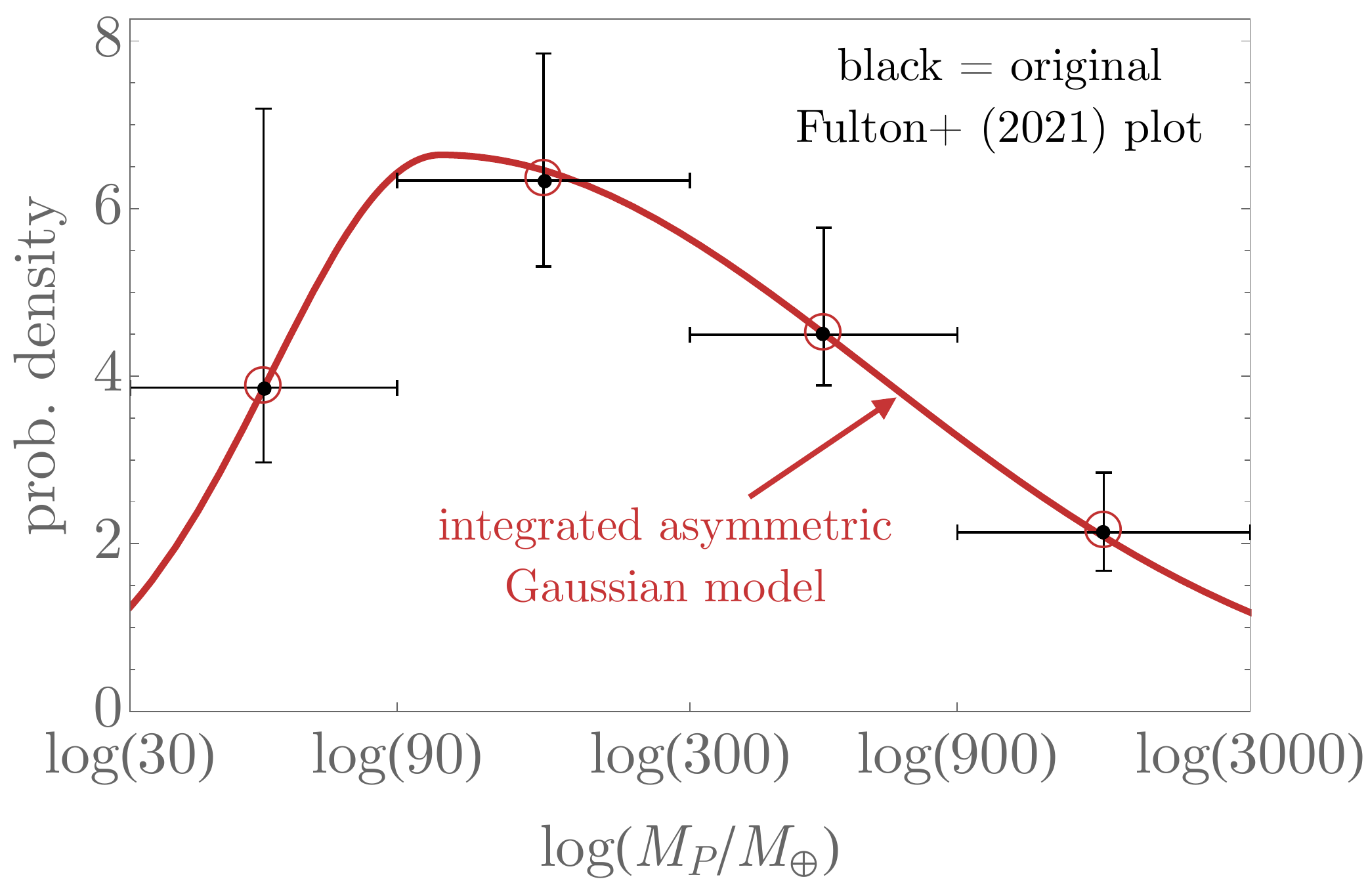}
\caption{
Comparison of the original data presented in Figure~1 of \citet{fulton:2021}
for the binned occurence rate of exoplanets in the range 1-5\,AU (black),
with our proposed asymmetric Gaussian model (red).
}
\label{fig:fulton}
\end{center}
\end{figure}

The asymmetric normal was integrated over the four bins of that figure,
and then we defined a likelihood function between our integrated model
occurrence rate model and the reported occurrence rates in each bin.
Since the reported uncertainties are asymmetric, we again use asymmetric
normals for the likelihood functions. We then maximized the likelihood
function with respect to $\mu$, $\sigma_{-}$, $\sigma_{+}$ and a $y$-axis
scaling parameter, $f$, obtaining $\mu=4.678$, $\sigma_{-}=0.697$,
$\sigma_{+}=1.791$ and $f=20.706$. A comparison of our integrated best
fitting model to the original \citet{fulton:2021} plot is shown in
Figure~\ref{fig:fulton}. We did not pursue an MCMC approach here to infer
posteriors for these terms, since with four data points and four unknowns
we are precariously degenerate already.

Equipped with this model, we draw a random variate from the unbinned
distribution, representing a fair value of $\log(M_P)$. Masses exceeding
14\,$M_J$ are rejected given the upper limit constraint from
\citet{vanderburg:2020}. The mass is then converted into a radius using
the \forecaster\ package \citep{chen:2017}, again rejected any radii outside
of our range of 0.1 to 20\,$R_{\oplus}$ for which we have computed the
completeness.

We note that the resulting radius distribution is radically different from
that found using \kepler\ statistics. If we integrate the \citet{he:2019}
radius distribution, the probability of obtaining a $R_P<2$\,$R_{\oplus}$
(potentially terrestrial) exoplanet is 83\%, or 75\% after trimming
tidally disrupted planets. In contrast, the \citet{fulton:2021} distribution
gives a probability of $R_P<2$\,$R_{\oplus}$ of just 0.016\%, both with
and without Roche limit trimming.
In other words, terrestial planets are incredibly rare in this alternative
model. This is surely a consequence of the assumed asymmetric Gaussian.
Since there is no data below 30\,$M_{\oplus}$ here, the model is simply
extrapolating and fundamentally assumes a monotonically decreasing function
away from the mean. In reality, the distribution could equally plateau, but
terrestrial $>$AU exoplanets are simply beyond the sensitivity of CLS or
indeed any other radial velocity survey (for now), and thus this situation
is somewhat unavoidable. Regardless, the point here isn't whether the assumed
distribution is truly valid for terrestrial planets, but rather it represents
an aggressively ``top-heavy'' distribution juxtaposing the ``bottom-heavy''
\kepler\ model.

\subsection{Evaluating the Surprisingness}

\begin{figure*}
\begin{center}
\includegraphics[width=18.0cm,angle=0,clip=true]{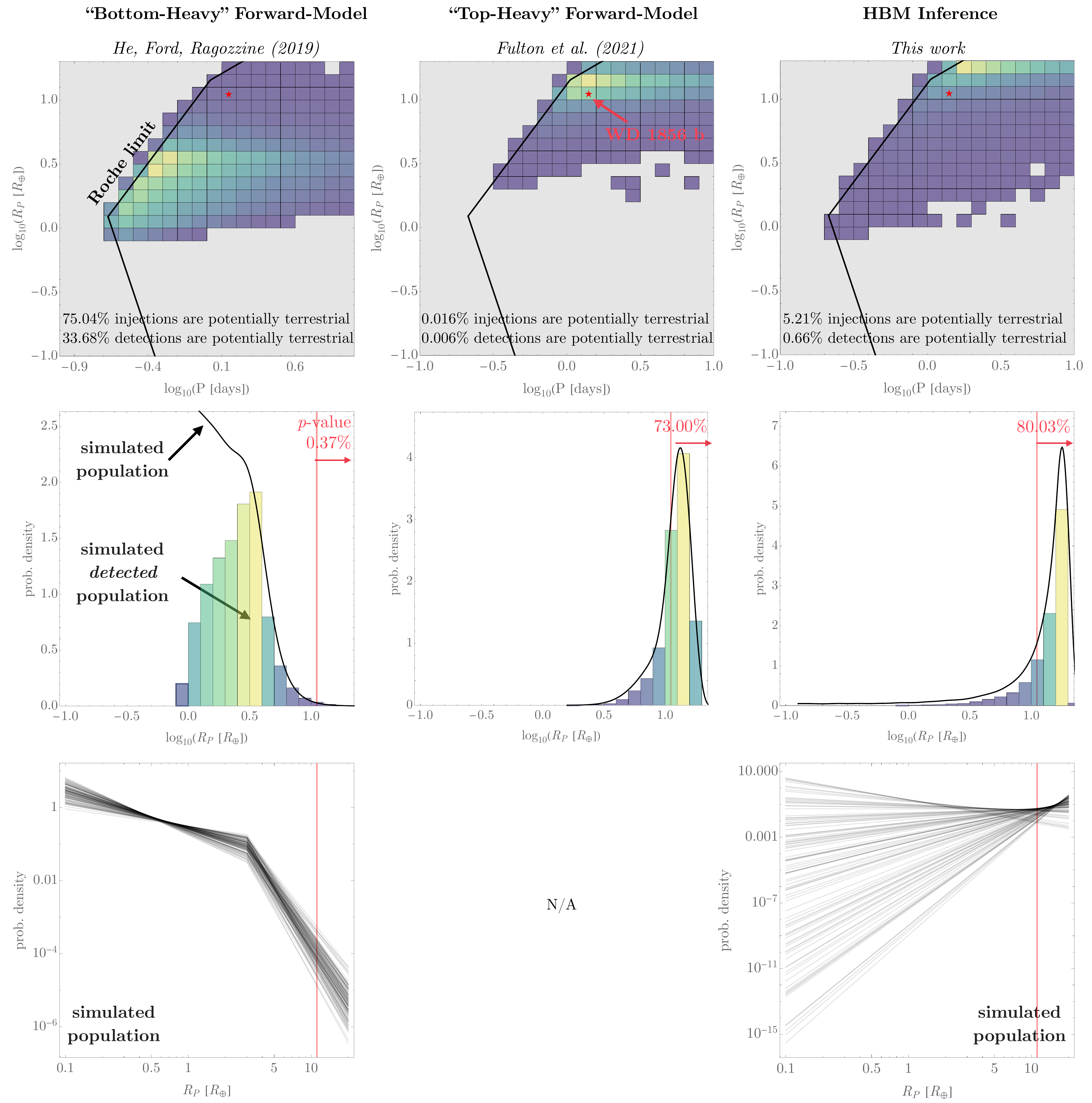}
\caption{
Three models (each column) for the radius distribution of WD exoplanets
and the resulting detection yields expected around a WD\,1856-twin star
with a sector of \tess\ data. The bottom-heavy \textit{Kepler}-inspired
model (left) produces a yield map (top row) that makes WD\,1856\,b an
outlier. In contrast, the the top-heavy model (middle) naturally explains
WD\,1856\,b but predicts very few terrestrial planets. The right column
shows a simple power-law inference using an HBM, which favours
to a top-heavy distribution. The middle row shows the marginalized radius
dimension of the upper row, and the bottom row shows the intrinsic
distributions which from they were drawn.
}
\label{fig:forward}
\end{center}
\end{figure*}

In each model, we randomly generated planets until $10^5$ transiting, detected
planets were found for a WD\,1856-twin star. The simulated detections define a
probability map of expected yield, and against that we can compare the location
of WD\,1856\,b. The results are summarized in Figure~\ref{fig:forward}.

Consider first the bottom-heavy \textit{Kepler}-inspired model. In this model,
any hopes we might have for Earth-sized planets around WDs are well founded.
They exist in abundance, comprising 75\% of the exoplanet population
and 34\% of the detected population. If this model holds, it would be
totally reasonable to expect JWST, LSST, etc to soon reveal pale blue dots
around these diminutive stars. However, in this model, the probability that the
first transiting exoplanet would have a radius of greater than or equal to that
of WD\,1856\,b is just 0.37\%. It is perfectly possible that
WD\,1856\,b is simply a rather improbable coincidence in this sense, but a much
simpler explanation is that the radius distribution of close-in WD exoplanets
does not resemble a bottom-heavy distribution, as assumed here. And of course,
such a conclusion would have profound implications for our ongoing searches.

Switching to the top-heavy model inspired from \citet{fulton:2021}, the
surprisingness of detected a WD\,1856\,b or larger planet is a completely
plausible 73\% now. However, this comes at a significant cost. The
probability of detecting a terrestrial planet is just 0.016\%, less than 1 in
6000. So this top-heavy model is consistent with the first WD transiting
exoplanet being a giant, but it projects a pessimistic forecast on the chance
of ever detecting terrestrial planets around WDs.

As noted earlier, a reasonable concern here with this top-heavy model is the
fact it extrapolates in mass. And indeed similarly a concern with the
bottom-heavy model is that it extrapolates in semi-major axis. Regardless,
both concur that terrestrial planets around WDs are not expected, given
WD\,1856\,b. As a final approach, we instead simply attempt to infer the
radius distribution from the single point of WD\,1856\,b directly in the
next section.

\section{Hierarchical Bayesian Inference}
\label{sec:hbm}

\subsection{Setting up the HBM}

In the previous section, we proposed two possible radius distributions for the
WD exoplanet population and evaluated the surprisingness of the first detected
transiting planet being a giant planet like WD\,1856\,b. However, whilst both
proposed distributions have physical motivations to justify them, neither can
be claimed to be representative since we have no previous population of WD
exoplanets from which to calibrate our expectations. Instead of proposing
a specific radius distribution and evaluating its surpringness, in this section
we infer the radius distribution conditioned upon WD\,1856\,b.

The parameters of WD\,1856\,b, $\boldsymbol{\theta}$, can be inferred from
transit modeling of the available photometry, as was done by
\citet{vanderburg:2020}. In principle, one could infer the parameters
describing the radius distribution, $\boldsymbol{\phi}$, conditioned upon
$\boldsymbol{\theta}$ with a simple Bayesian inference model. However, those
terms not only have uncertainty but also those uncertainties will be somewhat
informed by whatever radius distribution is assumed. This speaks to the need
for joint inference of both the underlying radius distribution and the planet
parameters, i.e. a hierarchical Bayesian model (HBM). In this framework, we
can write that

\begin{align}
\pdf(\boldsymbol{\theta},\boldsymbol{\phi}|\mathcal{D}) &=
\pdf(\mathcal{D}|\boldsymbol{\theta},\boldsymbol{\phi})
\pdf(\boldsymbol{\theta}|\boldsymbol{\phi}) \pdf(\boldsymbol{\phi})
\end{align}

The first term on the right-hand side is the likelihood function, the
probability of obtaining the data given some choice of input parameters. It's
easy to see that the likelihood here in our case is independent of the
radius distribution parameters, $\boldsymbol{\phi}$, and thus

\begin{align}
\pdf(\boldsymbol{\theta},\boldsymbol{\phi}|\mathcal{D}) &=
\pdf(\mathcal{D}|\boldsymbol{\theta})
\pdf(\boldsymbol{\theta}|\boldsymbol{\phi}) \pdf(\boldsymbol{\phi}).
\end{align}

We next note that our data, $\mathcal{D}$, is really three pieces of data.
First, we have the fact that the planet transits the star, denoted by symbol
$\hat{b}$. Second, we have the fact that this transiting planet was
detected by Sector 14 of \tess, denoted by $\hat{d}$. And third, we have
data concerning the transit shape, for which we'll retain the symbol
$\mathcal{D}$, thus giving us

\begin{align}
\pdf(\boldsymbol{\theta},\boldsymbol{\phi}|\hat{b},\hat{d},\mathcal{D}) &=
\pdf(\hat{b},\hat{d},\mathcal{D}|\boldsymbol{\theta})
\pdf(\boldsymbol{\theta}|\boldsymbol{\phi}) \pdf(\boldsymbol{\phi}).
\label{eqn:temp1}
\end{align}

We can now expand out the likelihood-like term in here, by
noting that $\pdf(A,B) = \pdf(A|B) \pdf(B)$:

\begin{align}
\pdf(\hat{b},\hat{d},\mathcal{D}|\boldsymbol{\theta}) &=
\pdf(\mathcal{D}|\hat{b},\hat{d},\boldsymbol{\theta}) \pdf(\hat{b},\hat{d}|\boldsymbol{\theta}),\nonumber\\
\qquad&= \pdf(\mathcal{D}|\boldsymbol{\theta}) \pdf(\hat{d}|\hat{b},\boldsymbol{\theta}) \pdf(\hat{b}|\boldsymbol{\theta}),
\end{align}

where the second line makes the notational simplification that if we have
$\mathcal{D}$ then conditions $\hat{b}$ and $\hat{d}$ are implicitly true.
Plugging this back into Equation~(\ref{eqn:temp1}), we have

\begin{align}
\pdf(\boldsymbol{\theta},\boldsymbol{\phi}|\hat{b},\hat{d},\mathcal{D}) &=
\pdf(\mathcal{D}|\boldsymbol{\theta}) \pdf(\hat{d}|\hat{b},\boldsymbol{\theta}) \pdf(\hat{b}|\boldsymbol{\theta})
\pdf(\boldsymbol{\theta}|\boldsymbol{\phi}) \pdf(\boldsymbol{\phi}).
\label{eqn:temp2}
\end{align}

For the transit data $\mathcal{D}$, we could use the full \tess\ light curve
but this would add significant computational cost to our inference and is
ultimately unnecessary since in our inference these parameters are ultimately
nuisance parameters we will marginalize over. Instead, we take refined transit
parameters from \citet{xu:2021} for this system, specifically we adopt their
white light curve inferred parameters from their Table~4 of $R_P/R_{\star}
\equiv p = (7.86\pm0.01)$ and $b=(7.75\pm0.01)$. We assume both quantities can
be described by normal distributions, such that

\begin{align}
\pdf(\mathcal{D}|\boldsymbol{\theta}) &= \frac{\exp(-\frac{(R_P/R_{\star}-\mu_p)^2}{2\sigma_p^2})}{\sqrt{2\pi}\sigma_p}
\frac{\exp(-\frac{(b-\mu_b)^2}{2\sigma_b^2})}{\sqrt{2\pi}\sigma_b},
\end{align}

where $\mu_p = 7.86$, $\sigma_p = 0.01$, $\mu_b = 7.75$ and $\sigma_b = 0.01$
\citep{xu:2021}. The term $\pdf(\hat{d}|\hat{b},\boldsymbol{\theta})$ describes
our completeness function, already described in detail in
Section~\ref{sec:completeness}. The next term,
$\pdf(\hat{b}|\boldsymbol{\theta})$, describes the transit probability, here
given by

\begin{align}
\pdf(\hat{b}|\boldsymbol{\theta}) &= \frac{ 1+R_P/R_{\star} }{ a/R_{\star} },\nonumber\\
\qquad&= \frac{ 1+R_P/R_{\star} }{ ^3\sqrt{\frac{\rho_{\star} G P^2}{3\pi}} },
\end{align}
	
where the second-line exploits Kepler's Third Law to remove the $a$ dependency
and instead use the mean stellar density for which we fix
$\rho_{\star} = 3.248\times10^8$\,kg\,m$^{-3}$ \citep{vanderburg:2020}.
Similarly, we fix $P =1.407939$\,days \citep{xu:2021} and $R_{\star} =
0.0131$\,$R_{\odot}$ \citep{vanderburg:2020}. The above implicitly assumes a
uniform prior on impact parameter, $b$.

We now need to choose $\pdf(\boldsymbol{\theta}|\boldsymbol{\phi})$, which
really means picking a parametric form for the radius distribution. Although
the parameters describing that distribution will be inferred
($\boldsymbol{\phi}$), a typical limitation of HBMs is that we must still
choose the mathematical form of that distribution. Given we only have a single
data point, we elected to use a power-law distribution, similar to that used
by \citet{howard:2012}. We did experiment with a broken power-law but found
the parameters could not converge. Accordingly, we have

\begin{align}
\pdf(\boldsymbol{\theta}|\boldsymbol{\phi}) &= \frac{
R_P^{-\alpha} (\alpha-1) }{
(R_{\mathrm{min}} R_{\mathrm{max}})^{-\alpha} (R_{\mathrm{min}} R_{\mathrm{max}}^{\alpha} - R_{\mathrm{min}}^{\alpha} R_{\mathrm{max}} )
},
\end{align}

where $\alpha$ is the power-law index and $R_{\mathrm{min}}$ \&
$R_{\mathrm{max}}$ are the minimum \& maximum radii allowed by the
distribution, for which we use the same 0.1$\,R_{\oplus}$ to
20$\,R_{\oplus}$ range over which our completeness model is defined.

Finally, for the hyper-prior, $\pdf(\boldsymbol{\phi}) = \pdf(\alpha)$, we
adopt an unbounded uniform prior such that in practice there's no need
to actually include this term.

\subsection{Results}

\begin{figure*}
\begin{center}
\includegraphics[width=18.0cm,angle=0,clip=true]{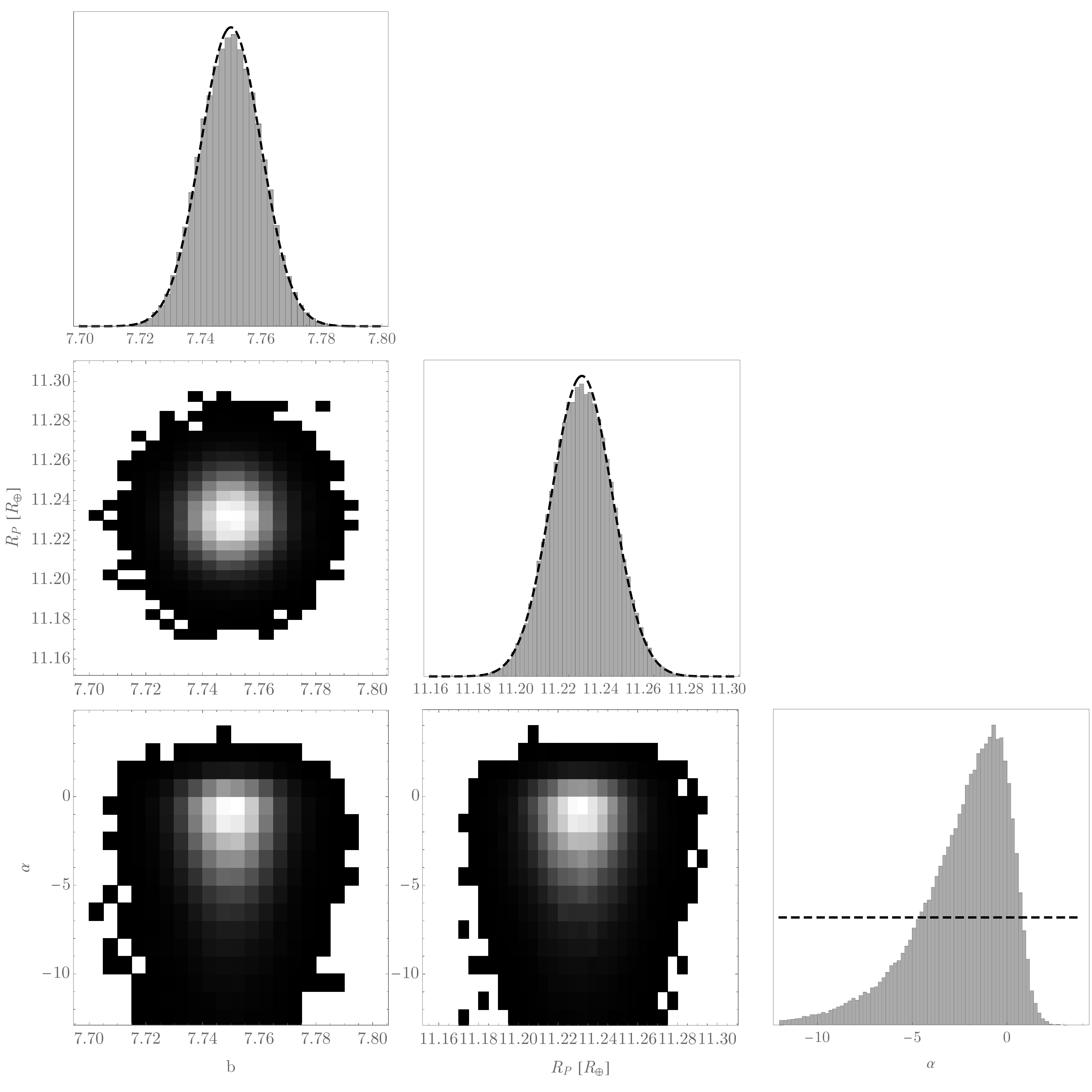}
\caption{
Corner plot of the posteriors from our HBM model. The parameters
($b$ and $R_P$) and hyper-parameters (just $\alpha$) are shown together
with the priors highlighted with the dashed lines on the marginalized
panels.
}
\label{fig:mcmc}
\end{center}
\end{figure*}

To infer both $\boldsymbol{\theta}$ and $\boldsymbol{\phi}$, which correspond
to the parameters $R_P$ \& $b$ and $\alpha$ respectively, we turn to a
Markov Chain Monte Carlo (MCMC) approach. We use a Metropolis sampler with
normal proposal functions using one million accepted steps. Burn-in points
were removed and mixing and convergence verified by inspection of the chains.
Finally, we thinned the resulting chain to 100,000 points. A corner plot of
the posteriors is presented in Figure~\ref{fig:mcmc}.

The posterior distributions for $R_P$ and $b$ closely match the strong prior
constraint we leverage from \citet{xu:2021}, as expected. After checking this
basic point, we turn to $\alpha$ - the power-law index for the radius
distribution. Since $\pdf(R_P|\alpha) \propto R^{-\alpha}$, then negative
indices correspond to top-heavy radius distributions (paucity of terrestrial
planets), whereas positive indices are bottom-heavy (abundance of terrestrial
planets). The median and $\pm34.15$\% quantile range of the marginalized
\textit{a-posteriori} distribution of $\alpha$
can be summarized as $\alpha = -1.9_{-2.9}^{+1.8}$.

From this value, one might naively conclude that there is a ${\sim}1$-$\sigma$
preference for a top-heavy radius distribution, which implies a less significant
discrepancy than that indicated by the summary statistics quoted in the inset
panels of Figure~\ref{fig:forward}. The confusion here stems from the fact that
a ``bottom-heavy'' distribution could be reasonably defined as beginning once
$\alpha>0$, but really $\alpha=0$ would lead to a median radius of
$10.1$\,$R_{\oplus}$ still (i.e. nearly a Jupiter radius). Truly what matters
is whether the samples of distribution is dominated by potentially terrestrial
planets, defined in this work as occurring at 2\,$R_{\oplus}$. The transition
from a sample dominated by terrestrials (``terrestrial-heavy'') versus
dominated by Neptunian and Jovian worlds (``terrestrial-light'') in fact occurs
at $\alpha=+0.9$. Comparing this to our MCMC posteriors, we find that only
2.8\% of samples correspond to $\alpha>0.9$ ($2.2$\,$\sigma$) and this is
a more accurate assessment of the HBM significance.

\subsection{Surprisingness of our Inferred Model}

As was done in Section~\ref{sec:pvalue}, we can take this model and ask
how surprising it would be for WD\,1856\,b to be the first detection.
Before attempting that calculation, we should expect the surprisingness to
be small, since the model is of course conditioned upon the detection
of WD\,1856\,b.

To accomplish this, we draw a random sample for $\alpha$ from our inferred
posterior distribution to create a radius distribution realization (100 such
realizations are shown in the bottom-right panel of Figure~\ref{fig:forward}).
We next draw a random radius from that distribution. As before, the period
distribution is assumed to be log-uniform from 0.1 to 10 days. We then draw a
random impact parameter for the planet between 0.01 and $a/R_{\star}$, where
the latter is indeed the maximum physically possible impact parameter possible
and the former is the smallest impact parameter over which our grid of
completeness models were trained (see Section~\ref{sec:completeness}).
Finally, we evaluate the detection probability using our model from
Section~\ref{sec:completeness}. If a planet is transiting and detected, the
resulting period, impact parameter and radius are saved are plausibly
detectable objects. This is repeated until $10^5$ such ``detections'' are
accumulated.

The resulting density of realized planets is shown in the top-right panel of
Figure~\ref{fig:forward}. From this, and the marginalized version shown in
the panel below, we can see that the vast majority of realizations are
consistent with the first planet being as large as WD\,1856\,b, indeed
80\% of realizations are such. However, the top-heavy nature of the
inferred distributions means that just 5.2\% of simulated planets are
potentially terrestrial ($<2$\,$R_{\odot}$).

This analysis supports the conclusions implied by Section~\ref{sec:pvalue};
the fact that WD\,1856\,b was the first detected transiting planet around
a WD implies that terrestrial planets are rare around them.

\newpage
\section{Discussion}
\label{sec:discussion}

\subsection{Summary}

To date, there is only one known transiting planet orbiting a white dwarf -
WD\,1856\,b \citep{vanderburg:2020}. The diminutive size of WDs has
long been recognized as a pathway to detecting Earth-sized or smaller planets
with existing facilities \citep{agol:2011}. Further, there is an emerging view
that Jupiter-sized planets represent the minority of the planet population
\citep{wittenmyer:2016,hsu:2018,fulton:2021}. Thus, the fact that the first
transiting planet detected around a WD was found to be a giant planet
is somewhat surprising.

We have quantified this surpringness for two possible radius distribution
models; one using the \textit{Kepler}-derived broken power-law model of
\citet{he:2019} and the other based off the radial velocity survey results
of \citet{fulton:2021}. The former is bottom-heavy (abundance of terrestrial
planets) and is in apparent contradiction with the detection of WD\,1856\,b,
with a $p$-value of $0.37$\%. However, WD\,1856\,b may have migrated
inwards during the post main-sequence evolution and thus the \kepler\
demographics (probing $<1$\,AU) are not representative. Using the
\citet{fulton:2021} mass distribution, converted into a radius distribution
using \forecaster\ \citep{chen:2017}, we find a top-heavy distribution more
consistent with WD\,1856\,b. However, such a distribution implies a dearth of
terrestrial planets, with just $0.016$\% of WDs possessing them in close-in
orbits.

Rather than propose speculative radius distributions, we also directly
fitted for a radius distribution using an HBM and again find that a top-heavy
distribution is indicated. Using a single-power law model, we find that
only 5\% of WDs would have $0.1$-$2$ Earth radii planets at
periods of 0.1 to 10\,days.

An implied conclusion from this work is that the great hope that transiting WD
planets might prove a shortcut to biosignature detection
\citep{kaltenegger:2020} may be a false hope, at least for exoplanets.
Specifically, the deduced hypothesis is that planets with radii
$0.1$-$2$\,$R_{\oplus}$ are much rarer (${\sim}$ an order of magnitude)
than $2$-$20$\,$R_{\oplus}$ around WDs for orbital periods
of $0.1$-$10$\,days (encompassing most WD habitable zones;
\citealt{agol:2011}). However, we note that exomoons of WD giant planets,
should they exist, would then potentially be a remaining source of optimism.

\subsection{Other Evidence for Exoplanets around WDs}

WD\,1856\,b may well be the only transiting exoplanet known, but other
exoplanet candidates have been reported around single WDs, but we note that all
of them are indeed Jupiter-mass or higher \citep{veras:2020}.
MOA-2010-BLG-477L\,b is a Jupiter-mass exoplanet found at ${\sim}3$\,AU and
detected via microlensing \citep{blackman:2021}.
At even an wider separation, WD\,0806-661\,b is a giant planet or sub-brown
dwarf at 2500\,AU detected via direct imaging \citep{luhman:2011}.
Both of these are clearly at much wider separations than we have been focussed
on in this paper, that is exoplanets close enough for plausible transit
probabilities - as well being within the habitable zone \citep{kilic:2013} to
enable biosignature reconnassiance \citep{kaltenegger:2020}.

Perhaps the most useful exoplanet for comparison, although it remains only
a candidate \citep{gaia:2023}, is \gaia\ DR3
2098419251579450880\,b\footnote{Another candidate, WD\,0141–675\,b, was also
presented in that \citet{gaia:2023} but has since been retracted 
(see \href{https://www.cosmos.esa.int/web/gaia/dr3-known-issues}{here}).}.
Discovered through astrometry in \gaia\ DR3, this is a ${\sim}34$\,$M_J$
candidate companion, orbiting at
$242$\,days. Thus, the candidate is highly unlikely to exhibit transits but
perhaps represents the tail-end of the distribution we are seeing in the
0.1-10\,days regime (where they are expected). In any case, it is
almost certainly roughly Jupiter-sized and thus is fully consistent with the
hypothesis proposed in this work. However, unlike the \tess\ data, the
\gaia\ data is not sensitive to terrestrial planets for these stars
\citep{perryman:2014} and thus we cannot state that its existence adds
greater weight to our argument, rather that it merely appear consistent with
it.

Concordantly, none of the other known exoplanet candidates around WDs really
speak to the claimed hypothesis. A more interesting observation is that many
WDs show evidence for metal contaminated atmospheres indicative of accreting
material \citep{koester:2014}. These observations have been interpreted as
likely have been caused by tidally disrupted rocky bodies \citep{jura:2008,
debes:2019} and thus would seem to imply the existence of close-in terrestrial
planets afterall. However, this doesn't necessarily contradict this work's
hypothesis, since the progenitor population could be one which, if scattered
inwards, is consistently scattered inwards sufficient to lead to tidal
disruption.

The case of WD\,1145+017 (discovered with K2 photometry) perhaps serves as a
possible example \citep{vanderburg:2015}, where an ensemble of co-orbiting
transits were detected around the host star at approximately the Roche limit,
indicative of fragments and dust from a planetary disruption. ZTF\,J0139+5245
is another such example discovered with ZTF \citep{vanderbosch:2020}, but a
more extreme one, since the period is far beyond the circular Roche limit
implying either a high eccentricity ($e>0.97$) or rotational fission
\citep{veras:2020}. Some additional possible candidate examples are reported
in \citet{guidry:2021}.

An alternative resolution is that the minor bodies indicated above are in fact
dominated by progenitor moons rather than planets, as been suggested previously
\citep{doyle:2021}.

\subsection{Other Possibilities}

We can conceive of two other possible ways to resolve the evidence for WD
minor bodies with the surprisingly large size of WD\,1856\,b. The first
is that our assumed single power-law is wrong in enforcing a monotonically
increasing probability density with respect to radius. Perhaps the
distribution turns over at some radius, representing the most unlikely
planetary radius, and then peaks back up. There's an infinite number of
possible such distributions and thus we haven't attempted to propose
specific speculations, but such a distribution would necessarily have to
turn over at around an Earth radius in order to be consistent with the
WD\,1856\,b first detection (else terrestrial planets would dominate
the detected population again).

A second possibility is that WD\,1856\,b is simply a fluke. Perhaps there
truly is a bottom-heavy distribution and it was indeed highly improbable
that a WD\,1856\,b-sized exoplanet would be the first to be revealed in
transit. Using the \textit{Kepler}-like bottom-heavy distribution, we found
the odds of this to be 0.37\% earlier in Section~\ref{sec:pvalue}.
That's certainly interesting, but hardly overwhelming - in the history of
astronomy, improbable events can and will occur given enough time.

For these reasons, we don't consider our hypothesis in any way established
with conviction. It would certainly be premature to abort on-going and 
future efforts to look for terrestrial planets around WDs. However, it is a
curious implication resulting from the discovery of WD\,1856\,b that deserves
further attention. Fortunately, the hypothesis is eminently testable through
on-going searches and thus we can look forward to a firmer resolution in the
future.
	
\section*{Acknowledgments}

This paper includes data collected by the TESS mission, which are publicly available from the Mikulski Archive for Space Telescopes (MAST).
We acknowledge the use of public TESS Alert data from pipelines at the TESS Science Office and at the TESS Science Processing Operations Center.
Funding for the TESS mission is provided by NASA’s Science Mission directorate.
This research has made use of the Exoplanet Follow-up Observation Program website, which is operated by the California Institute of Technology, under contract with the National Aeronautics and Space Administration under the Exoplanet Exploration Program.

Thanks to anonymous reviewer for their constructive report.
DK thanks
M. Sloan,
D. Daughaday,
A. Jones,
E. West,
T. Zajonc,
C. Wolfred,
L. Skov,
G. Benson,
A. de Vaal,
M. Elliott,
M. Forbes,
S. Lee
Z. Danielson,
C. Souter,
M. Gillette,
T. Jeffcoat,
J. Rockett,
T. Donkin,
A. Schoen,
J. Black,
R. Ramezankhani,
S. Marks,
N. Gebben,
M. Hedlund,
D. Bansal,
J. Sturm,
Rand Corp.,
L. Deacon,
R. Provost,
B. Sigurjonsson,
B. Walford,
N. De Haan,
J. Gillmer,
E. Garland,
A. Leishman,
Queen Rd. Fnd. Inc.,
B. Pearson,
S. Thayer,
B. Seeley,
F. Blood,
I. Williams,
J. Smallbon,
B. Sahlberg \&
X. Yao.

\section*{Data Availability}

We make the two simulated detected populations from Section~\ref{sec:pvalue}
publicly available, as well as the version for the HBM inferred model from
Section~\ref{sec:hbm} (all of which were presented in Figure~\ref{fig:forward})
at \wwwcoolworlds. Similarly, we make the HBM posteriors available in the same
repository (described in Section~\ref{sec:hbm} and presented in
Figure~\ref{fig:mcmc}) as well as our MCMC inference code. We also make the
code underlying the model shown in Figure~\ref{fig:fulton} available, as well
as the resulting simulated radius samples. Finally, the simulated Sector 14
SNRs are made available, as well as the corresponding training grid, and
the resulting detection contours calculated from them (as well as the code
to do).

\appendix

\bsp
\label{lastpage}

\begin{thebibliography}{99}
\bibitem[\protect\citeauthoryear{Agol}{2011}]{agol:2011} Agol E., 2011, ApJL, 731, L31. doi:10.1088/2041-8205/731/2/L31
\bibitem[\protect\citeauthoryear{Beatty \& Gaudi}{2008}]{beatty:2008} Beatty T.~G., Gaudi B.~S., 2008, ApJ, 686, 1302. doi:10.1086/591441
\bibitem[\protect\citeauthoryear{Blackman et al.}{2021}]{blackman:2021} Blackman J.~W., Beaulieu J.~P., Bennett D.~P., Danielski C., Alard C., Cole A.~A., Vandorou A., et al., 2021, Natur, 598, 272. doi:10.1038/s41586-021-03869-6
\bibitem[\protect\citeauthoryear{Bonsor et al.}{2020}]{bonsor:2020} Bonsor A., Carter P.~J., Hollands M., G{\"a}nsicke B.~T., Leinhardt Z., Harrison J.~H.~D., 2020, MNRAS, 492, 2683. doi:10.1093/mnras/stz3603
\bibitem[\protect\citeauthoryear{Borucki et al.}{2003}]{borucki:2003} Borucki W.~J., Koch D.~G., Lissauer J.~J., Basri G.~B., Caldwell J.~F., Cochran W.~D., Dunham E.~W., et al., 2003, SPIE, 4854, 129. doi:10.1117/12.460266
\bibitem[\protect\citeauthoryear{Burke et al.}{2015}]{burke:2015} Burke C.~J., Christiansen J.~L., Mullally F., Seader S., Huber D., Rowe J.~F., Coughlin J.~L., et al., 2015, ApJ, 809, 8. doi:10.1088/0004-637X/809/1/8
\bibitem[\protect\citeauthoryear{Chen \& Kipping}{2017}]{chen:2017} Chen J., Kipping D., 2017, ApJ, 834, 17. doi:10.3847/1538-4357/834/1/17
\bibitem[\protect\citeauthoryear{Christiansen et al.}{2016}]{christiansen:2016} Christiansen J.~L., Clarke B.~D., Burke C.~J., Jenkins J.~M., Bryson S.~T., Coughlin J.~L., Mullally F., et al., 2016, ApJ, 828, 99. doi:10.3847/0004-637X/828/2/99
\bibitem[\protect\citeauthoryear{Claret et al.}{2020}]{claret:2020} Claret A., Cukanovaite E., Burdge K., Tremblay P.-E., Parsons S., Marsh T.~R., 2020, A\&A, 634, A93. doi:10.1051/0004-6361/201937326
\bibitem[\protect\citeauthoryear{Cort{\'e}s \& Kipping}{2019}]{cortes:2019} Cort{\'e}s J., Kipping D., 2019, MNRAS, 488, 1695. doi:10.1093/mnras/stz1300
\bibitem[\protect\citeauthoryear{Dame et al.}{2019}]{dame:2019} Dame K., Belardi C., Kilic M., Rest A., Gianninas A., Barber S., Brown W.~R., 2019, MNRAS, 490, 1066. doi:10.1093/mnras/stz398
\bibitem[\protect\citeauthoryear{Danielski et al.}{2019}]{danielski:2019} Danielski C., Korol V., Tamanini N., Rossi E.~M., 2019, A\&A, 632, A113. doi:10.1051/0004-6361/201936729
\bibitem[\protect\citeauthoryear{Debes et al.}{2019}]{debes:2019} Debes J.~H., Th{\'e}venot M., Kuchner M.~J., Burgasser A.~J., Schneider A.~C., Meisner A.~M., Gagn{\'e} J., et al., 2019, ApJL, 872, L25. doi:10.3847/2041-8213/ab0426
\bibitem[\protect\citeauthoryear{Doyle, Desch, \& Young}{2021}]{doyle:2021} Doyle A.~E., Desch S.~J., Young E.~D., 2021, ApJL, 907, L35. doi:10.3847/2041-8213/abd9ba
\bibitem[\protect\citeauthoryear{Feroz, Hobson, \& Bridges}{2009}]{feroz:2009} Feroz F., Hobson M.~P., Bridges M., 2009, MNRAS, 398, 1601. doi:10.1111/j.1365-2966.2009.14548.x
\bibitem[\protect\citeauthoryear{Fleming et al.}{2020}]{fleming:2020} Fleming S.~W., Barclay T., Bell K.~J., Bianchi L., Brasseur C.~E., Hermes J., Loyd R.~O.~P., et al., 2020, arXiv, arXiv:2010.00007. doi:10.48550/arXiv.2010.00007
\bibitem[\protect\citeauthoryear{Foreman-Mackey, Hogg, \& Morton}{2014}]{dfm:2014} Foreman-Mackey D., Hogg D.~W., Morton T.~D., 2014, ApJ, 795, 64. doi:10.1088/0004-637X/795/1/64
\bibitem[\protect\citeauthoryear{Fressin et al.}{2013}]{fressin:2013} Fressin F., Torres G., Charbonneau D., Bryson S.~T., Christiansen J., Dressing C.~D., Jenkins J.~M., et al., 2013, ApJ, 766, 81. doi:10.1088/0004-637X/766/2/81
\bibitem[\protect\citeauthoryear{Fulton et al.}{2014}]{fulton:2014} Fulton B.~J., Tonry J.~L., Flewelling H., Burgett W.~S., Chambers K.~C., Hodapp K.~W., Huber M.~E., et al., 2014, ApJ, 796, 114. doi:10.1088/0004-637X/796/2/114
\bibitem[\protect\citeauthoryear{Fulton et al.}{2021}]{fulton:2021} Fulton B.~J., Rosenthal L.~J., Hirsch L.~A., Isaacson H., Howard A.~W., Dedrick C.~M., Sherstyuk I.~A., et al., 2021, ApJS, 255, 14. doi:10.3847/1538-4365/abfcc1
\bibitem[\protect\citeauthoryear{Gaia Collaboration et al.}{2023}]{gaia:2023} Gaia Collaboration, Arenou F., Babusiaux C., Barstow M.~A., Faigler S., Jorissen A., Kervella P., et al., 2023, A\&A, 674, A34. doi:10.1051/0004-6361/202243782
\bibitem[\protect\citeauthoryear{Guerrero et al.}{2021}]{guerrero:2021} Guerrero N.~M., Seager S., Huang C.~X., Vanderburg A., Garcia Soto A., Mireles I., Hesse K., et al., 2021, ApJS, 254, 39. doi:10.3847/1538-4365/abefe1
\bibitem[\protect\citeauthoryear{Guidry et al.}{2021}]{guidry:2021} Guidry J.~A., Vanderbosch Z.~P., Hermes J.~J., Barlow B.~N., Lopez I.~D., Boudreaux T.~M., Corcoran K.~A., et al., 2021, ApJ, 912, 125. doi:10.3847/1538-4357/abee68
\bibitem[\protect\citeauthoryear{He, Ford, \& Ragozzine}{2019}]{he:2019} He M.~Y., Ford E.~B., Ragozzine D., 2019, MNRAS, 490, 4575. doi:10.1093/mnras/stz2869
\bibitem[\protect\citeauthoryear{Howard et al.}{2010}]{howard:2010} Howard A.~W., Marcy G.~W., Johnson J.~A., Fischer D.~A., Wright J.~T., Isaacson H., Valenti J.~A., et al., 2010, Sci, 330, 653. doi:10.1126/science.1194854
\bibitem[\protect\citeauthoryear{Howard et al.}{2012}]{howard:2012} Howard A.~W., Marcy G.~W., Bryson S.~T., Jenkins J.~M., Rowe J.~F., Batalha N.~M., Borucki W.~J., et al., 2012, ApJS, 201, 15. doi:10.1088/0067-0049/201/2/15
\bibitem[\protect\citeauthoryear{Hsu et al.}{2018}]{hsu:2018} Hsu D.~C., Ford E.~B., Ragozzine D., Morehead R.~C., 2018, AJ, 155, 205. doi:10.3847/1538-3881/aab9a8
\bibitem[\protect\citeauthoryear{Jura}{2008}]{jura:2008} Jura M., 2008, AJ, 135, 1785. doi:10.1088/0004-6256/135/5/1785
\bibitem[\protect\citeauthoryear{Jura \& Young}{2014}]{jura:2014} Jura M., Young E.~D., 2014, AREPS, 42, 45. doi:10.1146/annurev-earth-060313-054740
\bibitem[\protect\citeauthoryear{Kaltenegger et al.}{2020}]{kaltenegger:2020} Kaltenegger L., MacDonald R.~J., Kozakis T., Lewis N.~K., Mamajek E.~E., McDowell J.~C., Vanderburg A., 2020, ApJL, 901, L1. doi:10.3847/2041-8213/aba9d3
\bibitem[\protect\citeauthoryear{Kang, Liu, \& Shao}{2021}]{kang:2021} Kang Y., Liu C., Shao L., 2021, AJ, 162, 247. doi:10.3847/1538-3881/ac23d8
\bibitem[\protect\citeauthoryear{Kilic et al.}{2013}]{kilic:2013} Kilic M., Agol E., Loeb A., Maoz D., Munn J.~A., Gianninas A., Canton P., et al., 2013, arXiv, arXiv:1309.0009. doi:10.48550/arXiv.1309.0009
\bibitem[\protect\citeauthoryear{Kipping}{2010}]{binning:2010} Kipping D.~M., 2010, MNRAS, 408, 1758. doi:10.1111/j.1365-2966.2010.17242.x
\bibitem[\protect\citeauthoryear{Kipping \& Tinetti}{2010}]{nightside:2010} Kipping D.~M., Tinetti G., 2010, MNRAS, 407, 2589. doi:10.1111/j.1365-2966.2010.17094.x
\bibitem[\protect\citeauthoryear{Kipping et al.}{2012}]{kipping:2012} Kipping D.~M., Bakos G. {\'A}., Buchhave L., Nesvorn{\'y} D., Schmitt A., 2012, ApJ, 750, 115. doi:10.1088/0004-637X/750/2/115
\bibitem[\protect\citeauthoryear{Kipping}{2023}]{kipping:2023} Kipping D., 2023, MNRAS, 523, 1182. doi:10.1093/mnras/stad1492
\bibitem[\protect\citeauthoryear{Koester, G{\"a}nsicke, \& Farihi}{2014}]{koester:2014} Koester D., G{\"a}nsicke B.~T., Farihi J., 2014, A\&A, 566, A34. doi:10.1051/0004-6361/201423691
\bibitem[\protect\citeauthoryear{Lagos et al.}{2021}]{lagos:2021} Lagos F., Schreiber M.~R., Zorotovic M., G{\"a}nsicke B.~T., Ronco M.~P., Hamers A.~S., 2021, MNRAS, 501, 676. doi:10.1093/mnras/staa3703
\bibitem[\protect\citeauthoryear{van Lieshout et al.}{2018}]{lieshout:2018} van Lieshout R., Kral Q., Charnoz S., Wyatt M.~C., Shannon A., 2018, MNRAS, 480, 2784. doi:10.1093/mnras/sty1271
\bibitem[\protect\citeauthoryear{Limbach et al.}{2022}]{limbach:2022} Limbach M.~A., Vanderburg A., Stevenson K.~B., Blouin S., Morley C., Lustig-Yaeger J., Soares-Furtado M., et al., 2022, MNRAS, 517, 2622. doi:10.1093/mnras/stac2823
\bibitem[\protect\citeauthoryear{Luhman, Burgasser, \& Bochanski}{2011}]{luhman:2011} Luhman K.~L., Burgasser A.~J., Bochanski J.~J., 2011, ApJL, 730, L9. doi:10.1088/2041-8205/730/1/L9
\bibitem[\protect\citeauthoryear{Mandel \& Agol}{2002}]{mandel:2002} Mandel K., Agol E., 2002, ApJL, 580, L171. doi:10.1086/345520
\bibitem[\protect\citeauthoryear{Mordasini et al.}{2012}]{mordasini:2012} Mordasini C., Alibert Y., Georgy C., Dittkrist K.-M., Klahr H., Henning T., 2012, A\&A, 547, A112. doi:10.1051/0004-6361/201118464
\bibitem[\protect\citeauthoryear{Morris et al.}{2021}]{morris:2021} Morris B.~M., Heng K., Brandeker A., Swan A., Lendl M., 2021, A\&A, 651, L12. doi:10.1051/0004-6361/202140913
\bibitem[\protect\citeauthoryear{Mustill et al.}{2018}]{mustill:2018} Mustill A.~J., Villaver E., Veras D., G{\"a}nsicke B.~T., Bonsor A., 2018, MNRAS, 476, 3939. doi:10.1093/mnras/sty446
\bibitem[\protect\citeauthoryear{Nordhaus \& Spiegel}{2013}]{nordhaus:2013} Nordhaus J., Spiegel D.~S., 2013, MNRAS, 432, 500. doi:10.1093/mnras/stt569
\bibitem[\protect\citeauthoryear{Perryman et al.}{2014}]{perryman:2014} Perryman M., Hartman J., Bakos G. {\'A}., Lindegren L., 2014, ApJ, 797, 14. doi:10.1088/0004-637X/797/1/14
\bibitem[\protect\citeauthoryear{Sanderson, Bonsor, \& Mustill}{2022}]{sanderson:2022} Sanderson H., Bonsor A., Mustill A., 2022, MNRAS, 517, 5835. doi:10.1093/mnras/stac2867
\bibitem[\protect\citeauthoryear{Stephan, Naoz, \& Zuckerman}{2017}]{stephan:2017} Stephan A.~P., Naoz S., Zuckerman B., 2017, ApJL, 844, L16. doi:10.3847/2041-8213/aa7cf3
\bibitem[\protect\citeauthoryear{Shvartzvald et al.}{2023}]{shvartzvald:2023} Shvartzvald Y., Waxman E., Gal-Yam A., Ofek E.~O., Ben-Ami S., Berge D., Kowalski M., et al., 2023, arXiv, arXiv:2304.14482. doi:10.48550/arXiv.2304.1448
\bibitem[\protect\citeauthoryear{Tamanini \& Danielski}{2019}]{tamanini:2019} Tamanini N., Danielski C., 2019, NatAs, 3, 858. doi:10.1038/s41550-019-0807-y2
\bibitem[\protect\citeauthoryear{van Sluijs \& Van Eylen}{2018}]{vansluijs:2018} van Sluijs L., Van Eylen V., 2018, MNRAS, 474, 4603. doi:10.1093/mnras/stx3068
\bibitem[\protect\citeauthoryear{Vanderbosch et al.}{2020}]{vanderbosch:2020} Vanderbosch Z., Hermes J.~J., Dennihy E., Dunlap B.~H., Izquierdo P., Tremblay P.-E., Cho P.~B., et al., 2020, ApJ, 897, 171. doi:10.3847/1538-4357/ab9649
\bibitem[\protect\citeauthoryear{Vanderburg et al.}{2015}]{vanderburg:2015} Vanderburg A., Johnson J.~A., Rappaport S., Bieryla A., Irwin J., Lewis J.~A., Kipping D., et al., 2015, Natur, 526, 546. doi:10.1038/nature15527
\bibitem[\protect\citeauthoryear{Vanderburg et al.}{2015}]{2015Natur.526..546V} Vanderburg A., Johnson J.~A., Rappaport S., Bieryla A., Irwin J., Lewis J.~A., Kipping D., et al., 2015, Natur, 526, 546. doi:10.1038/nature15527
\bibitem[\protect\citeauthoryear{Vanderburg et al.}{2020}]{vanderburg:2020} Vanderburg A., Rappaport S.~A., Xu S., Crossfield I.~J.~M., Becker J.~C., Gary B., Murgas F., et al., 2020, Natur, 585, 363. doi:10.1038/s41586-020-2713-y
\bibitem[\protect\citeauthoryear{Veras, McDonald, \& Makarov}{2020}]{veras:2020} Veras D., McDonald C.~H., Makarov V.~V., 2020, MNRAS, 492, 5291. doi:10.1093/mnras/staa243
\bibitem[\protect\citeauthoryear{Veras}{2021}]{veras:2021} Veras D., 2021, orel.book, 1. doi:10.1093/acrefore/9780190647926.013.238
\bibitem[\protect\citeauthoryear{Wittenmyer et al.}{2016}]{wittenmyer:2016} Wittenmyer R.~A., Butler R.~P., Tinney C.~G., Horner J., Carter B.~D., Wright D.~J., Jones H.~R.~A., et al., 2016, ApJ, 819, 28. doi:10.3847/0004-637X/819/1/28
\bibitem[\protect\citeauthoryear{Xu et al.}{2021}]{xu:2021} Xu S., Diamond-Lowe H., MacDonald R.~J., Vanderburg A., Blouin S., Dufour P., Gao P., et al., 2021, AJ, 162, 296. doi:10.3847/1538-3881/ac2d26

\end{thebibliography}
\end{document}